# Where are all the Sirius-Like Binary Systems?

J. B. Holberg[1*], T. D. Oswalt[2], E. M. Sion[3], M. A. Barstow[4] and M. R. Burleigh[4]

[1] *Lunar and Planetary Laboratory, Sonnett Space Sciences Bld., University of Arizona, Tucson, AZ 85721, USA*
[2] *Florida Institute of Technology, Melbourne, FL. 32091, USA*
[3] *Department of Astronomy and Astrophysics, Villanova University, 800 Lancaster Ave. Villanova
   University, Villanova, PA, 19085, USA*
[4] *Department of Physics and Astronomy, University of Leicester, University Road, Leicester LE1 7RH, UK*

**ABSTRACT**

Approximately 70% of the nearby white dwarfs appear to be single stars, with the remainder being members of binary or multiple star systems. The most numerous and most easily identifiable systems are those in which the main sequence companion is an M star, since even if the systems are unresolved the white dwarf either dominates or is at least competitive with the luminosity of the companion at optical wavelengths. Harder to identify are systems where the non-degenerate component has a spectral type earlier than M0 and the white dwarf becomes the less luminous component. Taking Sirius as the prototype, these latter systems are referred to here as 'Sirius-Like'. There are currently 98 known Sirius-Like systems. Studies of the local white dwarf population within 20 pc indicate that approximately 8 per cent of all white dwarfs are members of Sirius-Like systems, yet beyond 20 pc the frequency of known Sirius-Like systems declines to between 1 and 2 per cent, indicating that many more of these systems remain to be found. Estimates are provided for the local space density of Sirius-Like systems and their relative frequency among both the local white dwarf population and the local population of A to K main sequence stars. The great majority of currently unidentified Sirius-Like systems will likely turn out to be closely separated and unresolved binaries. Ways to observationally detect and study these systems are discussed.

[*] E-mail: holberg@argus.lpl.arizona.edu

## 1 INTRODUCTION

It was the irregular proper motions of several bright nearby stars that ultimately lead to the discovery of white dwarfs (WDs). The existence of the faint companions of Sirius and Procyon were first deduced from the unexpectedly perturbed proper motion noted in these stars by Bessel (1844). The companion to the A-type star Sirius was finally accidently found in 1862 as a curiously dim star, some ten magnitudes fainter than the primary. Thirty-four years later a similarly faint companion was observed orbiting the F-type star Procyon. Between 1913 and 1915 the odd spectral classification and the bluish color of the faint companion to the K-type star 40 Eridani respectively led Russell and Hertzsprung independently to recognize the subluminous nature of 40 Eri B (Holberg 2010). All of these systems lie within 5 pc of the Sun.

In this paper the term 'Sirius-Like System' (SLS) refers to WDs in binary or multiple star systems containing at least one component of spectral type K or earlier. These systems are typically distinguished from the much more numerous WD + M star systems by the fact that the WD is subordinate in luminosity and frequently in mass to their companions. Relatively hot WDs like Sirius B have absolute luminosities that match late K main sequence stars, meaning that they become increasingly difficult to find in close proximity to earlier main sequence spectral types. A key fortuitous advantage of some SLSs is that it is often possible to use them to investigate the WD initial-final mass relation (IFMR). The age of the current luminous star can be estimated from its luminosity and metallicity or activity, which gives the overall age of the system. Because the WD progenitor originally had a larger mass and earlier spectral type than the present non-degenerate component, it is often possible to gauge the original progenitor mass from its cooling age. The result is that the WD can be placed on the IFMR (see Liebert et al. 2005, Catalán et al. 2008a, Catalán et al. 2008b and Zhao et al. 2012). Extreme systems such as θ Hyd, where a massive WD closely orbits a B star, represent some of the best opportunities to identify the most massive main sequence stars able to form WDs, currently thought to be approximately 8 $M_\odot$. Subsequent stellar evolution of such primaries may lead to a type 1a supernova (Parthasarthy et al. 2007). SLSs also provide constraints on the mass and mass-ratio properties of main sequence stars in binary systems (Ferrario, 2012). Finally, SLSs are quite useful in efforts to empirically place the WD on the degenerate mass-radius relation MRR, (Holberg et al. 2012).

The primary modes of WD discovery have strongly influenced our perception of the relative frequency of different WD binary systems. For example, proper motion surveys yield numerous examples of WDs in common proper motion pairs, which biases discovery towards widely separated systems. Likewise ground-based blue and UV-excess surveys readily identify numerous close and unresolved white dwarf + red dwarf pairs, but fail to reveal WDs near more luminous companions. On the other hand space-based observations and surveys have identified a quite different set of SLSs characterized by very close or unresolved WDs orbiting luminous primary stars. These are primarily hot H-rich DA stars detected by UV or EUV excesses. These systems are thus heavily biased against cooler WDs and He-rich stars that often lack readily detectable short wavelength excesses.

In this paper the properties of known resolved and unresolved SLSs are presented. Included is the pre-cataclysmic variable system V471 Tau which is a DA + K2V system. Excluded is HS1136+6646 (Sing et al. 2004) which appears to be DAO.5 + K7V system but the estimated mass of the K7 star (0.34 $M_\odot$) indicates that it may actually be an irradiated M star. Also excluded are WD + subdwarf systems and currently interacting systems. Excluded as well are probable SLSs such as the single-line spectroscopic binary HD 209295 (Handler et al. 2002), which has a close 1 $M_\odot$ companion, but lacks the detection of an unambiguous WD photosphere.

In Section 2 the known SLSs are cataloged (Table 1) and discussed in terms of resolved and unresolved systems, which often reflects various discovery modes. Section 3 contains an examination of the inferred orbital properties of these systems. In Section 4 the observed spatial distribution of SLSs is discussed, which strongly suggests that many are missing from surveys beyond 20 pc. Space densities and the relative frequency of SLSs are also considered. Section 5 discusses how additional systems might be discovered and studied in the future.

## 2 KNOWN SIRIUS-LIKE SYSTEMS

Table 1 lists the 98 known Sirius-Like Systems, ordered by their Villanova white dwarf numbers (McCook & Sion 1998), along with an alternate designation (column 2). Column 3 gives a common designation for the luminous companion and the respective spectral types of the WD and companion in columns 4 and 5. Column 6 gives the effective temperature of the WD and columns 7 and 8 give the respective V magnitudes of the WD and companion. Columns 9 through 12 give the binary separation in arc seconds, the position angle, distance in pc and distance uncertainty, respectively. The orientation of the position angle is from the WD to the companion. The number of known components in each system is given in column 13.

### 2.1 Resolved systems

Approximately 70 per cent of the systems in Table 1 are resolved, including five (Sirius, Procyon, 40 Eri, HR 5692, and 56 Per) with fully determined orbits or where orbital motion has been observed astrometricaly (56 Per) and spectrometrically (HR 4343). Many of the resolved systems were discovered by traditional proper motion surveys, primarily those conducted by Willem Luyten. Most of the Luyten systems were discussed by Oswalt, Hintzen & Luyten (1988, hereafter OHL) for WDs with magnitudes brighter than 17 and by Oswalt & Strunk (1994, hereafter OS94) for WDs fainter than 17. The systems in the latter reference have been subject to very limited observational study. Angular separations for these wide systems range from approximately $10 - 10^2$ arcsec. Included among the resolved systems are nine very close ($\rho < 2.0"$) systems that have now been resolved with *HST* (see below).

### 2.2 Unresolved systems

Contemporary space-based surveys, primarily the *Extreme Ultraviolet Explorer (EUVE)* and the *ROSAT* Wide Field Camera (WFC), as well as the *International Ultraviolet Explorer (IUE)* have revealed many previously unresolved SLSs. The circumstances surrounding these discoveries are discussed in Barstow & Holberg (2003). The existence of these WDs was only apparent from the pronounced short wavelength spectroscopic or photometric excesses with respect to the main sequence stars. For many of the SLS WDs observed with *IUE,* good quality flux-corrected and coadded low dispersion spectra[1] can be found in Holberg, Barstow & Burleigh (2002). Approximately 30 per cent of currently known SLSs are of this type. These latter discoveries are strongly biased towards hot ($T_{eff} > 16,000$ K) DA WDs. Largely overlooked by space-based observations are cooler WDs having weak UV and EUV luminosities and non-DA WDs where photospheric opacities due to He and heavier elements effectively block observable EUV emission. One significant exception is WD 1643-573, which was discovered to be a hot He-rich DOZ companion to the G9 V star HD 149499 from an objective prism image of a star field obtained during the 1973 *Skylab* Mission (Parsons et al. 1976). Not only is WD 1643-573 a UV-excess object but it is hot enough and near enough that there is also observable EUV emission between 250 Å to 300 Å (Jordan et al. 1997). As mentioned previously the systems showing short wavelength excesses are heavily biased against cooler WDs and He-rich stars. Nevertheless, the existing sample of SLSs containing hot DA stars was very efficiently surveyed out to several hundred pc by the *ROSAT WFC* and *EUVE* all sky surveys and can be considered relatively complete, unless the WD companions contain significant levels of photospheric heavy element opacity or are heavily obscured by interstellar photoelectric absorption.

---

[1] Fully reduced coadded spectra can be found at: http://vega.lpl.arizona.edu/newsips/low/

A major improvement to knowledge of the orbital parameters of the unresolved UV and EUV SLSs resulted from a 'snapshot' program utilizing the *Hubble Space Telescope* (*HST*) *Wide Field Planetary Camera* 2 (*WFPC2*) images. Near UV filters were used to balance the flux between the WD and non-degenerate components. Barstow et al. (2001, hereafter B01) succeeded in resolving 8 out of 17 observed targets. (A subsequent observation of an 18[th] target also resulted in resolving a WD + G4V pair). For these resolved systems apparent separations ranged from 0.217″ to 2.0″. One system, 56 Per (WD 0421+338), was discovered to consist of four components; the previously recognized components 56 Per A and B were each seen to be double. The WD 56 Per Ab was resolved at a separation of 0.390″ from the F4V primary. In a second multiple system 14 Aur, (WD 0512+326, see appendix A) B01 resolved the WD and clarified the relation between the various components. For the nine unresolved systems the separations were presumed to be less than 0.08″, the FWHM of the *WFPC2* images. Table 1 includes the separations and position angles reported by B01. A follow-on *HST* Cycle 20 effort has obtained Balmer line spectra of a number of systems resolved by B01.

Table 1.  Known Sirius-Like Systems

| WD | Alt. ID | Primary | Type | Type | WD$_{Teff}$ | WD(V) | P(V) | ρ(″) | θ(°) | D(pc) | σ | Comp. |
|---|---|---|---|---|---|---|---|---|---|---|---|---|
| (1) | (2) | (3) | (4) | (5) | (6) | (7) | (8) | (9) | (10) | (11) | (12) | (13) |
| 0022-745 | HD 2133 B | HD 2133 | DA1.9 | F7V | 26800 | 15.6 | 9.62 | 0.602 | 46.85 | 136.61 | 17.36 | 2 |
| 0023-109 | G 158-078 | G 158-077 | DA4.8 | K6V | 10430 | 16.22 | 12.9 | 59.58 | 150.34 | 68 | 4 | 3 |
| 0030+444 | G 172-4 | BD +43 100 | DA4.8 | K1V | 10480 | 16.44 | 10.28 | 29.12 | 31.54 | 89.13 | 12.07 | 2 |
| 0040-220 | SDSS | SDSS | WD | K7.1V | ... | 17.79 | 18.34 | 22.655 | 228.18 | 800 | ... | 2 |
| 0041+092 | BD +08 102 B | BD +8 102 | DA2.1 | K2V | 23740 | 14.2 | 10.04 | <0.08 | ... | 65 | ... | 2 |
| 0042+140 | LP 466-033 | BD +13 99 | DC | G8V | ... | 19.19 | 9.79 | 71.77 | 103.6 | 69.54 | 6.96 | 2 |
| 0052+115 | LP 466-421 | LP 466-420 | DB | G | ... | 18.3 | 16.8 | 7.12 | 270.8 | ... | ... | 2 |
| 0114+027 | AY Cet B | HD 7672 | DA2.8 | G5III | 18000 | ... | 5.43 | unres | ... | 80.58 | 1.75 | 2 |
| 0208-510 | GJ 86 B | GJ 86 | DQ6 | K0V | 8180 | 13.2 | 6.17 | 1.93 | 103.4 | 10.78 | 0.04 | 3 |
| 0210+086 | ξ$^1$ Cet B | HD 13611 | DA3.9 | G8Iab | 13000 | ... | 4.37 | <0.16 | ... | 117.51 | 7 | 2 |
| 0221+399 | NLTT 7890 | BD +39 0539 | DA8.1 | K3V | 6250 | 17.39 | 8.94 | 40.74 | 157.51 | 38.0 | 3 | 2 |
| 0226-615 | HD 15638 B | HD 15638 | DA1.1 | F6V | 47840 | 14.7 | 8.86 | <0.08 | ... | 206.19 | 33.16 | 2 |
| 0227-005 | LP 590-190 | BD -01 343 | DB4.4 | K0V | 12702 | 19.07 | 9.95 | 10.75 | 24.67 | 65 | ... | 2 |
| 0250-007 | LP 591-177 | BD -1 407 | DA6.0 | G5V | 8400 | 16.4 | 9.11 | 27.41 | 32.48 | 37 | 3 | 2 |
| 0252-055 | HD 18131 B | HD 18131 | DA2.0 | K0IV | 30540 | 14.5 | 7.32 | <0.08 | ... | 101.83 | 9.75 | 2 |
| 0304+154 | LP 471-52 | LP 471-51 | DC | K3V | ... | 20.3 | 11.48 | 18.3 | 188.2 | 71.4 | ... | 2 |
| 0315-011 | LP 592-80 | BD -1 469 | DA6.7 | K1IV | 7520 | 17.2 | 5.37 | 48.43 | 48.066 | 67.16 | 3.79 | 2 |
| 0338-388 | HE 0338-3853 | ... | DA.63 | F2IV/V | 80000 | 16.1 | 10.5 | unres | ... | 240 | ... | 2 |
| 0347+171 | V471 Tau B | V471 Tau | DA1.5 | K2V | 34200 | 13.64 | 9.48 | orbit | ... | 44.07 | 2.93 | 3 |
| 0353+284 | ... | V1092 Tau | DA1.5 | K2V | 32948 | ... | 11.51 | unres | ... | 105 | ... | 2 |
| 0354-368 | EUVE J | MS0354.6-3650 | DA1 | G2V | 53000 | 17.4 | 12.45 | 0.992 | 73.44 | 400 | ... | 2 |
| 0400-346 | NLTT 12412 | HD 25535 | DA9.9 | G1V | 5100 | 17.82 | 6.8 | 64.15 | 312.83 | 51.68 | 1.68 | 2 |
| 0413-077 | 40 Eri B | HD 26965 | DA3.1 | K0.5V | 16400 | 9.53 | 4.41 | 83.42 | 102.99 | 4.98 | 0.01 | 3 |
| 0415-594 | ε Ret B | HD 27442 | DA3.3 | K2IV | 15310 | 12.5 | 4.44 | 12.8 | ... | 18.24 | 0.05 | 3 |
| 0418+137 | HD 27483 B | HD 27483 | DA3.1 | F6V | 16420 | 14.5 | 6.14 | 1.276 | 10.62 | 47.42 | 1.15 | 3 |
| 0421+338 | 56 Per Ab | HD 27786 | DA3.1 | F4V | 16500 | 15 | 5.8 | 0.39 | 306.29 | 40.62 | 0.79 | 4 |
| 0433+270 | G 039-027 | HD 283750 | DA9 | K7V | 5620 | 15.6 | 8.42 | 123.9 | 338.86 | 17.97 | 0.46 | 2 |
| 0457-103 | HD 32008 B | HR 1608 | DA2.0 | K0IV | 24910 | 13.1 | 5.4 | <0.08 | ... | 53.97 | 2.45 | 2 |
| 0458-364 | ... | RE J0500-362 | DA1.1 | F6V | 45000 | 17.9 | 13.7 | unres | ... | 530 | ... | 2 |
| 0512+326 | 14 Aur Cb | 14 Aur Ca | DA1.3 | F2V | 39740 | 14.1 | 7.88 | 2.0 | 290.11 | 103.84 | 31.48 | 5 |
| 0551+123 | G 105-2 | HD 39570 | DB3.8 | F8V | 13200 | 15.83 | 7.76 | 90.12 | 231.82 | 53.53 | 2.32 | 2 |
| 0551+560 | LP 120-26 | LP 120-27 | DC | G9V | ... | 18.1 | 12.81 | 8.54 | 122.59 | ... | ... | 2 |
| 0615-591 | L 182-061 | HD 44120 | DB3.2 | F9.5V | 15570 | 14.03 | 6.43 | 40.98 | 301.65 | 37.43 | 0.41 | 2 |
| 0642-166 | Sirius B | Sirius | DA2 | A0V | 24790 | 8.44 | -1.47 | 7.5 | ... | 2.63 | 0.01 | 2 |
| 0642-285 | LP 895-41 | CD -28 3358 | DA5.4 | K3V | 9370 | ... | 10.57 | 15.89 | 73.04 | 32 | 2 | 2 |
| 0658+712 | LP 34-137 | BD +71 380 | DC | G2V | ... | 19.8 | 6.238 | 30 | 260 | 81.5 | 9.1 | 2 |
| 0659+130 | ... | RE J0702+129 | DA1.4 | K0IV/V | 36910 | ... | 10 | unres | ... | 115 | ... | 3 |
| 0727-387 | y Pup B | HD 59635 | DA1.3 | B5Vp | 37600 | 14.52 | 5.39 | unres | ... | 165 | ... | 2 |
| 0736+053 | Procyon B | Procyon | DQZ6.5 | F5IV/V | 7740 | ... | 0.39 | orbit | ... | 3.51 | 0.02 | 2 |
| 0743-336 | VB 03 | HR 3018 | DC11.0 | G0V | 4600 | 16.59 | 5.37 | 869.65 | 2.81 | 15.21 | 0.12 | 2 |
| 0834+576 | ... | SBSS0834+576 | DA | G | ... | ... | 16.5 | 17.36 | ... | ... | ... | 2 |
| 0842+490 | HD 74389 B | BD +49 1766 | DA1.3 | A2V | 39500 | 14.62 | 7.48 | 20.11 | 84.28 | 111.48 | 7.08 | 3 |
| 0845-188 | LP 786-6 | BD -18 2482 | DB2.9 | K3V | 17470 | 15.682 | 11.33 | 31 | ... | 282 | ... | 2 |
| 0905-724 | HD 78791 B | HD 78791 | DA1.6 | F9II | 32500 | 14.5 | 4.48 | <0.08 | ... | 122.85 | 2.26 | 2 |
| 0911+023 | HR 3665 B | θ Hyd | DA1.6 | B9.5V | 30700 | ... | 3.88 | unres | ... | 34.8 | 1.88 | 3 |
| 0930+815 | HD 81817 B | HD 81817 | DA2.8 | K3III | 18000 | ... | 4.27 | unres | ... | 304.88 | 13.94 | 2 |
| 1004+665 | LP 62-35 | LP 62-34 | DZ | K4V | ... | 14.71 | 14.71 | 111.81 | 336.94 | 63 | ... | 2 |
| 1009-184 | LHS 2033 B | BD -17 3088 | DZ5.1 | K7V | 9940 | 15.44 | 9.91 | 399.55 | 192.91 | 17.18 | 0.49 | 2 |
| 1021+266 | HD 90052 B | BD +27 1888 | DA1.4 | F0V | 35040 | 15 | 9.3 | <0.08 | ... | 250 | ... | 2 |
| 1024+326 | ... | 2RE J1027+322 | DA1.4 | G5V | 41354 | ... | 13.2 | unres | ... | 380 | ... | 2 |
| 1027-039 | LP 670-9 | BD -3 2935 | DQ | GV | ... | 19.09 | 10.09 | unres | ... | 35.3 | ... | 2 |
| 1107-257 | LP 849-059 | HD 96941 | DC | G9V | ... | 16.79 | 8.69 | 100.02 | 179.72 | 40.16 | 1.58 | 2 |
| 1109-225 | β Crt B | β Crt | DA1.4 | A1III | 36740 | 13.4 | 4.61 | <0.08 | ... | 104.28 | 7.17 | 2 |
| 1130+189 | LP 433-6 | LP 433-7 | DA4.5 | G2V | 11200 | 17.65 | 14.54 | 154.25 | 169.95 | 66 | 5 | 2 |
| 1132-325 | VB 04 | HD 100623 | DC8 | K0V | 6300 | 16.7 | 5.98 | 16.04 | 129.3 | 9.56 | 0.04 | 2 |
| 1133+619 | LP 94-65 | LP 94-66 | DZ | K2V | ... | 18.37 | 12.31 | 17.637 | 67.29 | ... | ... | 2 |
| 1209-060 | LP 674-029 | HD 106092 | DA7.6 | K4V | 6180 | 17.26 | 9.96 | 203.28 | 102.31 | 45.19 | 3.04 | 2 |
| 1227+292 | LP 321-799 | LP 321-798 | DA2.2 | G2V | 22700 | 18.11 | 13.64 | 10.794 | 41.91 | 168 | 10 | 2 |
| 1250-226 | ... | BD -22 9659 | DAO.63 | G8IV | 80000 | ... | 9.69 | 0.11 | ... | 119.33 | 22.35 | 2 |
| 1304+227 | LP 378-537 | BD +23 2539 | DA4.8 | K0V | 10460 | 16.47 | 9.71 | 20.3 | 105.1 | 48 | 4.5 | 2 |
| 1306+083 | ... | RE J1309+081 | DA2 | G2V | >22000 | 16 | 11.5 | <0.08 | ... | 275 | ... | 2 |
| 1354+340 | G 165-B5B | BD +34 2473 | DA3.5 | F8V | 14490 | 16.16 | 9.08 | 54.82 | 114.25 | 99.4 | 11.02 | 2 |
| 1425+540 | G 200-39 | NLTT 37441 | DBA3.5 | K | 14490 | 15.04 | 15.04 | 60.146 | 172.37 | 57.8 | 13.03 | 2 |
| 1440+068 | LP 560-73 | HD 129517 | DZ | F8V | ... | 18.31 | 7.31 | 118.96 | 100.57 | 46 | ... | 2 |
| 1455+300 | LP 326-84 | BD +30 2592 | DA | K0IV | ... | 20.16 | 9.73 | 22.946 | 303.23 | 60.57 | 6.09 | 2 |
| 1501+301 | LP 326-74 | LP 326-75 | DC | K4V | ... | 17.84 | 15.61 | 87.996 | 114.82 | 79 | 5 | 2 |
| 1514+411 | LP 222-64 | LP 222-62 | DA6 | F | ... | 16.99 | 14.22 | 125.52 | 252 | 139 | 3 | 2 |
| 1516+207 | HD 136138 B | HR 5692 | DA1.7 | G8II-III | 30400 | 15.3 | 5.69 | orbit | ... | 98.8 | 6.5 | 2 |

| 1542+729 | LP 42-164 | LP 42-163 | DA | K0V | ... | 18.06 | 10.85 | 18.046 | 178.74 | 62.1 | 9.6 | 2 |
|---|---|---|---|---|---|---|---|---|---|---|---|---|
| 1544-377 | NLTT 41169 | HD 140901 | DA4.7 | G6V | 10820 | 12.8 | 6.01 | 15.2 | 130.52 | 15.35 | 0.09 | 2 |
| 1619+123 | PG 1629+123 | HD 147528 | DA2.9 | G0V | 17150 | 14.66 | 8.21 | 62.1 | 311.39 | 51.84 | 2.74 | 2 |
| 1620-391 | CD -38 10980 | HR 6094 | DA12.1 | G5V | 24406 | 11.00 | 5.38 | 345.08 | 247.58 | 12.78 | 0.06 | 3 |
| 1623+022 | LHS 3195 | BD +2 3101 | DC | K2V | ... | 18.03 | 10.74 | 9.3 | 319.53 | 56.69 | 6.81 | 2 |
| 1634-573 | HD 149499 B | HD 149499 | DOZ1 | G9V | 49500 | 11.7 | 8.7 | 1.5 | 58.5 | 36.4 | 1.56 | 2 |
| 1635+530 | 16 Dra B | HD 150100 | DA1.6 | B9.5V | 32000 | ... | 5.52 | unres | ... | 130.89 | 2.91 | 4 |
| 1659-531 | LP 0268-092 | HR 6314 | DA3.2 | F5V | 15570 | 13.47 | 5.29 | 113.76 | 257.76 | 27.23 | 0.47 | 2 |
| 1706+332 | G 181-B5B | BD +33 2834 | DA3.9 | F8V | 12960 | 15.92 | 8.59 | 35.18 | 84.49 | 69.69 | 4.22 | 2 |
| 1710+683 | LP 70-172 | LP 70-171 | DA7.6 | K7V | 6630 | 17.47 | 11.46 | 27.83 | 91.35 | 55 | 4 | 2 |
| 1730-370 | ... | λ Sco | DA.79 | B2IV | 64000 | ... | 1.62 | unres | ... | 175.13 | 23 | 2 |
| 1734+742 | 29 Dra B | HD 160538 | DA1.7 | K0III | 30000 | ... | 6.64 | unres | ... | 103.52 | 7.93 | 2 |
| 1736+133 | HD 160365 B | HD 160365 | DA2.2 | F6III | 23000 | ... | 6.14 | 8 | 180 | 99.9 | 4.79 | 2 |
| 1750+098 | G 140-B1B | HD 162867 | DC5.3 | K0V | 9527 | 15.75 | 9.36 | 24.78 | 206.74 | 49 | ... | 2 |
| 1803-482 | V 832 Ara B | HD 165141 | DA1.4 | G8III | 35000 | ... | 7.08 | unres | ... | 194.93 | 26.97 | 2 |
| 1848+688 | LDS 2421 B | BD +68 1027 | DA | G5V | ... | 17.18 | 9.72 | 34.59 | 274.43 | 78.86 | 4.73 | 2 |
| 1921-566 | CD -56 7708B | RE J1925-566 | DA1.0 | G5V | 49460 | 14.7 | 10.6 | 0.217 | 120.94 | 118.2 | 2.9 | 2 |
| 2048+809 | LP 25-436 | BD +80 670 | DA6 | G5V | 8440 | 16.59 | 9.08 | 18.78 | 321.75 | 40 | 3 | 2 |
| 2110+300 | ζ Cyg B | HD 202109 | DA4.2 | G8IIIp | 12000 | 13.2 | 3.2 | 0.036 | 246 | 43.88 | 0.67 | 2 |
| 2123-226 | ζ Cap B | HD 204075 | DA2.2 | G8IIIp | 23000 | ... | 3.75 | unres | ... | 118.2 | 2.93 | 2 |
| 2124+191 | HR 8210 B | HR 8210 | DA1.5 | A8V | 34320 | 14.4 | 6.08 | <0.08 | ... | 46.36 | 1.2 | 2 |
| 2129+000 | LP 638-004 | BD -00 4234 | DB3.5 | K2V | 14380 | 14.674 | 9.89 | 133.1 | 29.09 | 45.19 | 4.1 | 2 |
| 2217+211 | NLTT 53526 | BD +20 5125 | DC9 | K5V | 5600 | 17.69 | 10.07 | 83.04 | 283.27 | 49.26 | 3.4 | 2 |
| 2253-081 | NLTT 55288 | BD -8 5980 | DA7.4 | G6V | 6770 | 16.48 | 8.03 | 41.83 | 189.89 | 36.74 | 1.51 | 2 |
| 2257-073 | HD 217411 B | BD -7 5906 | DA1.4 | G5V | 36680 | 5600 | 9.88 | 1.01 | ... | 52.5 | ... | 2 |
| 2258+406 | G 216-B14B | G 216-B14A | DA5.1 | K9V | 9860 | 15.5 | 11.57 | 23 | 261 | 38.9 | 3.53 | 2 |
| 2301+762 | LP 027-275 | HD 218028 | DC6 | G5V | ~8400 | 16.35 | 8.64 | 12.898 | 139.29 | 66.8 | 3.53 | 2 |
| 2304+251 | 56 Peg B | HD 218356 | DA1 | K0Iab | 32000 | ... | 4.77 | unres | ... | 181.49 | 7.58 | 2 |
| 2350-083 | G 273-B1B | G 273-B1A | DA2.9 | GV | 17537 | 16 | 11 | 36 | 210 | 125 | 6 | 2 |
| 2350-706 | HD 223816 B | HD 223816 | DA.66 | G0V | 76200 | 14.04 | 9.89 | 0.574 | 37.02 | 116 | ... | 2 |

## 2.3 The barium stars

Among the SLSs there are a number of barium and mild barium stars, which have or are suspected of having WD companions. These stars are spectroscopically characterized by overabundances of Ba and other s-process elements. It is hypothesized that these over-abundances arose when the now WD companion was in its red giant phase and transferred s-process rich material to the now barium star. Under this hypothesis SLSs containing Ba stars are likely to have relatively close unresolved WD companions. It has long been noted that many Ba stars are themselves post main sequence stars (subgiants, giants and super giants) and radial velocity variables. This has led to the suggestion that all Ba stars have WD companions (Böhm-Vitense et al. 2000). Indeed, Böhm-Vitense et al. have investigated this issue by using *HST* STIS and Goddard High Resolution Spectrometer (GHRS) observations to search for the characteristic UV continuum of a WD shortward of the photospheric continuum in a number of Ba stars. They found convincing evidence of excess UV continua in five of their targets. Because these observed UV continua are weak and blended with chromospheric features from the Ba star none of these stars are included among the spectroscopically confirmed SLSs listed here but are included in a list of possible SLSs given in Table B1. Also included is WD 0353+248 (Jefferies & Smalley 1996) which is responsible for the EUV excess in the K2V primary, which has a barium excess. Recently Gray et al. (2011) using *GALEX* observations found photometric evidence for UV excess in six other Ba dwarf stars. This study helps to account for the putative WDs associated with dwarf Ba stars, which ought to exist in addition to the well known giants. As with the Böhm-Vitense et al. (2000) UV excesses, these recent Ba dwarf stars are listed as possible SLSs in Appendix B, Table B1. Among the Ba stars that are listed in Table 1 are ξ$^1$ Cet (WD 0210+086), 29 Dra (WD 1734+742), ζ Cyg (WD 2110+300) and ζ Cap (WD 2123-226).

From the above discussions it is clear that the sample of SLSs presented here is inherently heterogeneous, and constructed from distinctively different sources. Nevertheless, it is instructive to consider the distribution of spectral types and luminosity classes for both the WDs and Ba stars. For the WD stars the distribution in $T_{eff}$ and spectral type more closely resembles that of blue or UV excess magnitude-limited surveys such as the Palomar-Green (Green, Schmidt, & Liebert 1986) or Kiso Schmidt ultraviolet excess survey (Kondo, Noguchi, & Maehara 1984) surveys, than more inclusive volume-limited studies of local samples such as the 20 pc sample of Holberg et al. (2008). In particular, the SLSs contain some of the hottest and most luminous WDs as well as some of the coolest and least luminous. This is primarily due to the inclusion of systems characterized by UV and EUV excesses which emphasizes hot degenerate stars. For example, the most luminous stars in the local 20 pc WD sample have $T_{eff} \approx 25{,}000$ K, while Table 1 contains stars with $T_{eff}$ from 5000 K to well in excess of 70,000 K. For the non-degenerate components of Table 1 the distribution of spectral types runs from late K to early B, with the majority being dwarf stars. Many of the evolved luminosity classes correspond to barium stars with UV excesses.

## 3 ORBITAL PROPERTIES

Orbital or partial orbital elements are available for only a few of the WDs in Table 1. It is useful, however, to estimate basic properties of the remaining orbits. First, as with most binary systems, mass ratios, orbital radii and periods set constraints on the origin and evolution of such systems. Second, reasonable estimates of expected astrometric 'signals' which could lead to the discovery of new candidate SLSs are useful, therefore astrometric and radial velocity signatures have been computed in instances where existing information warrants. In general little is known about the actual orbits of these systems in Table 1. In such cases circular orbits have been assumed.

Table 2 contains the basic orbital information inferred for each system. For convenience Column 1 provides the same set of WD numbers as Table 1. Columns 2 and 3 give the respective mass estimates for the WD ($M_{WD}$) and companion ($M_C$) in solar masses, while Columns 4 and 5 provide their projected physical separations ($a_p$) in AU and the estimated semimajor axis ($a$) in AU, respectively. In the case of unresolved systems where there only exists an upper limit on the angular separation, these systems have assigned a corresponding upper limit to $a$. Where possible we have estimated masses for the WD and companion. For the WDs this includes dynamical masses from partial or complete orbits, or spectroscopic masses. For companions on the main sequence mass estimates are based on the empirical Mass-Luminosity relation of Xia & Fu (2010) using observed V magnitudes and distances from Table 1. For post main sequence companions no masses are estimated, unless they exist in the literature.

Column 6 gives either the measured or estimated orbital periods ($P$, in yrs). If the periods are known, either from an orbital solution or from radial velocity observations, they are denoted by italics. When the periods and masses are known the corresponding physical separations are computed from Kepler's third law and denoted by italics, except in the cases where complete orbital solutions exist. Column 7 gives the estimated orbital motion for resolved WD – Luminous star pairs in arc seconds per year. This is a useful quantity when considering potential uncertainties due to orbital motion for gravitational redshift measurements or high precision astrometry (see Section 5.1).

### 3.1 Resolved systems

The data in Table 1 was used to estimate the orbital characteristics given in Table 2. The projected physical separation $a_p$ in AU given in Column 3 is calculated from the product of the distance $D$ (in pc) and apparent angular separation $\rho$ (in arcseconds).

$$a_p = D\rho \tag{1}$$

In order to determine true semimajor axis ($a$), information on the eccentricity and orbital orientation is necessary, but it is often unavailable. Thus, some estimate of an average correction factor is necessary that effectively converts projected separations into semimajor axes. Dupuy & Liu (2011) have investigated this problem using Monte Carlo simulations to compute probability distributions of conversion factors that have been averaged over a range of time, eccentricity and orbital orientations and for various discovery biases. From Table 6 of Dupuy & Liu the case of circular orbits with moderate discovery biases was selected and a conversion factor of 1.11 is adopted; $a = 1.11a_p$. This only holds where the orbital period is unknown. For those unresolved systems where the orbital periods are known, the corresponding semimajor axes are computed from Kepler's third law and noted in italics. Using semimajor ($a$) and stellar masses estimated from Table 2, it is possible to estimate the corresponding orbital periods ($P$) in years from Kepler's third law. The semimajor axis is given in Column 5 of Table 2.

$$P = \sqrt{\frac{a^3}{(M_{WD}+M_C)}} \quad , \qquad (2)$$

For resolved systems the annual expected mutual orbital motion ($\Delta\theta$) of the WD-primary pair can be estimated from the mean motion ($2\pi/P$), the angular separation $\rho$ and the average value of the unknown orbital inclination, ($<\cos i> = 0.637$). This is given in Column 7 in units of µas as per year. Essentially this is an estimate of the expected change in the relative angular displacement between the components which might potentially be observable with extremely high precision astrometry.

$$\Delta\theta = 2\pi \langle \cos(i) \rangle \frac{10^6}{P} \quad , \qquad (3)$$

**Table 2. Orbital Properties**

|  | $M_\odot$ | $M_\odot$ | AU | AU | yrs | µas yr$^{-1}$ | µas yr$^{-2}$ | m s$^{-1}$yr$^1$ |  |
|---|---|---|---|---|---|---|---|---|---|
| 0022-745 | 0.53 | 1.19 | 82.2 | 91.3 | 665.03 | 3620.9 | 7.2 | 7.57E+00 |  |
| 0023-109 | 0.61 | 0.61 | 4051.4 | 4497.1 | 2.73E+05 | 872.9 | 4.23E-03 | 3.59E-03 |  |
| 0030+444 | 0.77 | 0.90 | 2595.5 | 2881.0 | 1.20E+05 | 973.4 | 1.08E-02 | 1.10E-02 |  |
| 0040-220 | … | 0.61 | 18124 | 20117.6 | … | … | … | … |  |
| 0041+092 | 0.39 | 0.84 | <5.2 | <5.8 | … | … | … | … |  |
| 0042+140 | 0.6 | … | 4990.9 | 5539.9 | … | … | … | … |  |
| 0052+115 | … | … | … | … | … | … | … | … |  |
| 0114+027 | 0.62 | 2.2 | … | *0.407* | *0.155* | 1.31E+05 | 1.11E+06 | 4.43E+05 | Appendix A |
| 0208-510 | 0.59 | 0.80 | 20.8 | 23.1 | 94.13 | 8.20E+04 | 1152.2 | 1.32E+02 | Appendix A |
| 0210+086 | 0.80 | 2.00 | <18.8 | *3.8* | *4.496* | *2.87E+04* | 8452.9 | *6.39E+03* | Barium star |
| 0221+399 | 0.79 | 0.83 | 1548.1 | 1718.4 | 5.60E+04 | 2911.7 | 0.069 | 3.18E-02 |  |
| 0226-615 | 0.54 | 1.70 | <16.5 | <18.3 | <52.3 | … | >492 | >1920 |  |
| 0227-005 |  | 1.64 | 698.8 | 775.6 | … | … | … | … |  |
| 0250-007 | 0.87 | 0.81 | 1014.2 | 1125.7 | 2.91E+04 | 3762.5 | 0.171 | 8.17E-02 |  |
| 0252-055 | 0.40 | … | <8.1 | <9.0 | … | … | … | … |  |
| 0304+154 | … | 0.71 | 1306.6 | 1450.3 | … | … | … | … |  |
| 0315-011 | 0.60 | … | 3252.6 | 3610.3 | … | … | … | … |  |
| 0338-388 | … | 1.27 | … | … | … | … | … | … |  |
| 0347+171 | 0.84 | 0.93 | … | *1.52E-02* | *1.42E-03* | *1.9E+09* | *1.8E+12* | *4.26E+08* | V471 Tau |
| 0353+284 | 0.6 | 0.79 | … | … | … | … | … | … | Barium star |
| 0354-368 | 0.7 | 1.08 | 396.8 | 440.4 | 6928.38 | 572.7 | 0.109 | 4.29E-01 |  |
| 0400-346 | 0.32 | 1.37 | 3315.3 | 3680.0 | 1.72E+05 | 1494.3 | 0.012 | 2.81E-03 |  |
| 0413-077 | 0.51 | 0.83 | 415.4 | 461.1 | *8000* | 4.17E+04 | 6.90 | 3.26E-01 | 40 Eri B |
| 0415-594 | 0.60 | 1.54 | 233.5 | 259.2 | *286* | 1.79E+05 | 827.82 | 1.06E+02 | Appendix A |
| 0418+137 | 0.98 | 3.04 | 60.5 | 67.2 | 274.53 | 1.86E+04 | 89.56 | 2.59E+01 | Appendix A |
| 0421+338 | 0.903 | 1.53 | 15.8 | 17.6 | 47.27 | 3.30E+04 | 923.17 | 3.48E+02 | Appendix A |
| 0433+270 | 0.62 | 0.72 | 2226.5 | 2471.4 | 1.06E+05 | 4669.5 | 0.058 | 1.21E-02 | |
| 0457-103 | 0.35 | .... | <4.3 | <4.8 | *2.47* | … | … | … | |
| 0458-364 | 0.68 | 0.97 | … | … | … | … | … | … | |
| 0512+326 | 0.56 | 1.51 | 207.7 | 230.5 | 2432.72 | 3.29E+03 | 1.787735 | 1.25E+00 | Appendix A |
| 0551+123 | 0.91 | 1.15 | 4824.1 | 5354.8 | 2.73E+05 | 1320.4 | 0.006396 | 3.78E-03 | |
| 0551+560 | | 0.85 | … | … | … | … | … | … | |
| 0615-591 | 0.61 | 1.28 | 1533.9 | 1702.6 | 5.11E+04 | 3207.7 | 0.083013 | 2.50E-02 | |
| 0642-166 | 1.0 | 2.03 | *19.7* | *21.9* | *50.00* | *6.00E+05* | *15870* | *3.44E+02* | Sirius |
| 0642-285 | 0.69 | 0.85 | 508.5 | 564.4 | 1.08E+04 | 5882.3 | 0.719962 | 2.58E-01 | |
| 0658+712 | … | 1.96 | 2445 | 2714.0 | … | … | … | … | |
| 0659+130 | 0.57 | .... | … | … | … | … | … | … | |
| 0727-387 | 1.0 | 5.17 | … | … | … | … | … | … | Appendix A |
| 0736+053 | 0.553 | 1.50 | … | *14.99* | *40.82* | *4.19E+05* | *13559.36* | *2.89E+02* | Procyon |
| 0743-336 | 0.59 | 1.08 | 13227.4 | 14682.4 | 1.38E+06 | 2526.8 | 0.002427 | 3.26E-04 | |
| 0834+576 | … | … | … | .... | … | … | … | … | |
| 0842+490 | 0.69 | 1.71 | 2241.9 | 2488.5 | 8.01E+04 | 1003.9 | 0.016568 | 1.33E-02 | Appendix A |
| 0845-188 | 0.58 | 1.16 | 8742 | 9703.6 | 7.25E+05 | 171.1 | 0.000312 | 7.33E-04 | |
| 0905-724 | 0.74 | | <9.8 | <10.9 | … | … | … | … | |
| 0911+023 | 1.21 | 2.52 | … | … | … | … | … | … | |
| 0930+815 | … | … | … | … | … | … | … | … | |
| 1004+665 | … | 0.54 | 7044.0 | 7818.9 | … | … | … | … | |
| 1009-184 | 0.59 | 0.61 | 6864.3 | 7619.3 | 6.07E+05 | 2632.4 | 0.005734 | 1.21E-03 | Appendix A |
| 1021+266 | 0.45 | 1.69 | <20 | <22.2 | … | … | … | … | |
| 1024+326 | 0.45 | 0.93 | … | … | … | … | … | … | |
| 1027-039 | … | 0.70 | … | … | … | … | … | … | |
| 1107-257 | … | 0.88 | 4016.8 | 4458.7 | … | … | … | … | |
| 1109-225 | 0.43 | … | <8.3 | <9.3 | *6* | … | … | … | Appendix A |
| 1130+189 | 0.95 | 0.55 | 10180.5 | 11300.4 | 9.81E+05 | 629.1 | 0.000848 | 8.85E-04 | |
| 1132-325 | … | 0.83 | 153.34 | 170.2 | … | … | … | … | Appendix A |
| 1133+619 | … | 0.74 | … | … | … | … | … | … | |
| 1209-060 | 0.60 | 0.76 | 9165.9 | 10174.1 | 8.80E+05 | 924.0 | 0.001389 | 6.90E-04 | |
| 1227+292 | 0.60 | 0.69 | 1813.4 | 2012.9 | 7.95E+04 | 543.0 | 0.009032 | 1.76E-02 | |
| 1250-226 | … | … | 13.1 | 14.6 | … | … | … | … | Appendix A |
| 1304+227 | 0.89 | 0.80 | 974.4 | 1081.6 | 2.74E+04 | 2967.6 | 0.143436 | 9.05E-02 | |
| 1306+083 | … | 1.11 | <22 | <24.4 | … | … | … | … | |
| 1354+340 | 0.65 | 1.16 | 5449.1 | 6048.5 | 3.50E+05 | 627.1 | 0.002372 | 2.11E-03 | |
| 1425+540 | 0.56 | 0.54 | 3476.4 | 3858.8 | 2.29E+05 | 1052.6 | 0.006091 | 4.48E-03 | |
| 1440+068 | … | 1.18 | 5472.2 | 6074.1 | … | … | … | … | |
| 1455+300 | 0.58 | … | 1389.8 | 1542.7 | … | … | … | … | |
| 1501+301 | 0.79 | 0.54 | 6951.7 | 7716.4 | 5.88E+05 | 598.9 | 0.001348 | 1.58E-03 | Appendix A |
| 1514+411 | 0.58 | 0.62 | 17446.6 | 19365.7 | 2.46E+06 | 204.1 | 1.1E-04 | 1.84E-04 | |
| 1516+207 | 0.80 | 1.84 | … | 1.98 | *1.71* | 4.23E+04 | 32744.78 | 2.44E+04 | Appendix A |
| 1542+729 | … | 0.74 | 1120.7 | 1243.9 | … | … | … | … | |
| 1544-377 | 0.72 | 0.97 | 233.3 | 259.0 | 3.21E+03 | 1.90E+04 | 7.82278 | 1.28E+00 | |
| 1619+123 | 0.55 | 1.05 | 3219.3 | 3573.4 | 1.69E+05 | 1470.9 | 0.011519 | 5.13E-03 | |
| 1620-391 | 0.68 | 1.01 | 4410.1 | 4895.2 | 2.63E+05 | 5239.2 | 0.026299 | 3.38E-03 | Appendix A |
| 1623+022 | … | 0.73 | 527.22 | 585.2 | … | … | … | … | |
| 1634-573 | 0.65 | 0.85 | 54.6 | 60.6 | 385.24 | 15574.8 | 53.46768 | 2.11E+01 | Appendix A |
| 1635+530 | 0.71 | 3.98 | … | … | … | … | … | … | |
| 1659-531 | 0.65 | 1.41 | 3097.7 | 3438.4 | 1.40E+05 | 3239.2 | 0.030495 | 6.54E-03 | |
| 1706+332 | 0.54 | 1.10 | 2451.7 | 2721.4 | 1.11E+05 | 1269.4 | 0.015144 | 8.68E-03 | |
| 1710+683 | 0.53 | 0.67 | 1530.7 | 1699.0 | 6.39E+04 | 1741.3 | 0.036021 | 2.18E-02 | |
| 1730-370 | 1.25 | 10.50 | … | 0.146 | 0.0163 | … | … | … | |
| 1734+742 | 0.65 | … | … | … | 14.24 | … | … | … | Barium star |
| 1736+133 | … | … | 799.2 | 887.1 | … | … | … | … | |
| 1750+098 | 1.17 | 0.85 | 1214.2 | 1347.8 | 3.48E+04 | 2847.1 | 0.108155 | 7.66E-02 | |
| 1803-482 | 0.66 | | … | … | 2.47 | … | … | … | Barium star |
| 1848+688 | … | 0.94 | 2727.8 | 3027.8 | … | … | … | … | |
| 1921-566 | 0.70 | 0.94 | 25.6 | 28.5 | 118.63 | 7317.1 | 81.5753 | 1.03E+02 | |
| 2048+809 | 0.81 | 0.84 | 751.2 | 833.8 | 1.87E+04 | 4007.6 | 0.282748 | 1.39E-01 | |
| 2110+300 | … | … | 1.6 | 1.8 | 17.8 | | 601.0605 | | Barium star |
| 2123-226 | … | … | … | … | … | … | … | … | Barium star |
| 2124+191 | 1.20 | 1.53 | <3.7 | 0.21 | 0.0595 | … | … | … | |
| 2129+000 | 0.49 | 0.77 | 6014.8 | 6676.4 | 4.86E+05 | 1095.5 | 0.002981 | 1.31E-03 | |
| 2217+211 | … | 0.77 | 4090.6 | 4540.5 | … | … | … | … | |
| 2253-081 | 0.52 | 0.95 | 1536.8 | 1705.9 | 5.81E+04 | 2879.3 | 0.065526 | 2.13E-02 | |
| 2257-073 | 0.49 | 0.95 | 53.0 | 58.9 | 376.29 | 1.07E+04 | 37.73363 | 1.68E+01 | |
| 2258+406 | 0.75 | 0.62 | 894.7 | 993.1 | 2.67E+04 | 3440.7 | 0.2 | 9.05E-02 | |
| 2301+762 | … | 1.07 | 861.6 | 956.4 | … | … | … | … | |
| 2304+251 | … | … | … | … | 0.3043 | | | | Barium star |
| 2350-083 | 0.57 | … | 4500 | 4995.0 | … | … | … | … | |
| 2350-706 | 0.67 | … | 66.6 | 73.9 | … | … | … | … | |

### 3.2 Very close and unresolved systems

For systems which are very close (those resolved by *HST*) or which remain unresolved it is possible to compute two additional quantities which can be potentially detected by the *Gaia* Mission or high precision radial velocity studies.

In the remainder of Table 2 Columns 8 and 9 provide the expected annual accelerations in angular velocity and radial velocity of the luminous component in $\mu$as yr$^{-2}$ and m s$^{-1}$ yr$^{-1}$, respectively. These quantities are useful in discussions of how additional SLSs might be recognized in future astrometric and radial velocity observations. Column 10 contains notes on particular systems. Unless actual orbits are known the quantities in Columns 8 and 9, assume circular orbits and averaged orbital inclinations. The relations used to estimate orbital dynamical values are given below and refer to the reflex motion of the primary star about the system centre of gravity. Since in most SLSs the luminosity of the primary star dominates the WD, the photocentre of the system very closely coincides with the physical centre of the primary star. Thus all that is observed is an angular acceleration in the star's proper motion given by

$$\dot{\mu} = 2\pi \rho_s \langle 0.842 \rangle \frac{10^6}{P^2} \qquad (4)$$

where $\rho_S = \rho M_{WD}/(M_{WD} + M_C)$. The numerical factor in the brackets results from averaging apparent circular orbits over an orbital period and averaging over inclination angles.

Column 9 contains an estimate of the expected radial velocity acceleration of the primary star and a measure of the expected radial velocity signal in unresolved systems (eqns. 5 and 6).

$$\dot{V} = (2\pi/P)^2 a_S \langle \sin i \rangle , \qquad (5)$$

where $a_S = aM_{WD}/(M_{WD} + M_C)$, and $\langle \sin i \rangle = 0.637$. The radial velocity acceleration (in units of m s$^{-1}$ yr$^{-1}$) which is appropriate for ongoing radial velocity studies that search for exoplanets is given by,

$$\dot{V} = 1.19 \times 10^5 a_S / P^2 . \qquad (6)$$

### 3.3 Mass Ratios and distribution of spectral types

Due to the limited range of WD masses and the lower mass limit of K stars - the mass ratio for SLSs, $q = M_{WD}/M_C$, vary from approximately 0.12 to 1.7 with a mean value of 0.68. Likewise, the mean mass of the WDs for which masses can be estimated is 0.68 M$_\odot$. This mass is approximately 0.09 M$_\odot$ larger than the spectroscopic masses of field WDs (Kleinman et al. 2013), which is not surprising given the fact that WDs in SLSs are the evolutionary descendents stars initially more massive than their current companions (see below).

The main sequence primary stars range from K9V to B5V with the following numbers by spectral type; B:3, A:3, F:14, G:27 and K:31. In Fig. 1 these numbers are compared with the corresponding relative space densities from Phillips et al. (2010). Although the numbers are modest (79 stars) it is evident that the two distributions show significant differences, in particular for G and K stars. This is perhaps not too surprising since the population of the SLSs is derived from both resolved CPM systems and largely unresolved UV and EUV excess systems. There are 20 post main sequence stars, sub giants, giants and super giants among the primary stars-many of these being barium stars.

### 4 SPACE DISTRIBUTION AND FREQUENCY

Here the present set of *known* SLSs is considered to be a reasonable approximation to the population of all Sirius-Like systems within a few hundred parsecs of the Sun. Figure 2 is a plot of the number of known SLSs in successive spherical shells of equal volume surrounding the Sun. There are 11 known SLSs[2] listed in Table 1 within 20 pc of the Sun, a spherical volume of $3.35 \times 10^4$ pc$^3$. Taking these to be representative of the true frequency of such systems, and presuming no further such systems will be found within 20 pc, the corresponding space density is $3.3 \times 10^{-4}$ pc$^{-3}$. With respect to the total number of all WDs within this volume (Sion et al. 2009, Holberg et al. 2008) this represents a relative frequency of approximately 8 per cent. It is also possible to estimate the volume-limited frequency of SLSs among the local population of main sequence stars from A to F. Phillips et al. (2010) determined volume-limited local stellar densities for main sequence stars over this range of spectral types. Summing over their individual space densities for each spectral type gives $5.71 \times 10^{-2}$ pc$^{-3}$. The ratio of this value to the space density of SLSs provides an estimate of ~ 0.6 percent as the frequency of SLSs among local main sequence stars. An alternate estimate of main sequence (B to K) space densities from the Kroupa, Tout & Gilmore (1993) initial mass function gives a density of $2.80 \times 10^{-2}$ pc$^{-3}$ and a corresponding frequency of SLSs among main sequence stars of 1.2 per cent. Thus, between 0.6 per cent and 1.2 per cent of all main sequence stars in the solar vicinity are expected to have WD companions. Among the 11 systems within 20 pc the distribution among spectral types is A:1, F:1, G:4, and K:5.

As is evident in Figure 2, the next four shells of equal volume contain *only one* additional SLS (WD1659-531 at 27 pc). Likewise, the average number of systems per shell, out to a distance of 50 pc, is only 1.5. There exist two possible additions to the number of SLSs just beyond 20 pc. These include Regulus (α Leo), a B7V star lying at a distance of 24.2 ± 0.2 pc which Gies et al. (2008) report to be a single-line radial velocity variable. They attribute this to a low mass WD companion with an orbital period of 40.11 days. We do not include this putative WD among our SLSs because there has been no detection of the companion's photosphere at any wavelength and the case cannot be conclusively made that the companion need be a WD. However, Regulus is included in Appendix B among the probable SLSs. If the companion is a WD at an orbital radius of 0.35 AU, it is not likely to reveal itself any time soon. There is also a potential CPM system containing a WD at 22.7 pc noted by Tokovinin & Lépine (2012) associated with the K0III star HD 197989. They identify GJ 9707C as a probable WD based on photometry however, we recently obtained a spectrum of this star that shows it is not a WD.

---

[2] WD 0208-510, WD 0413-077, WD 0415-594, WD 0433+270, WD 0642-166, WD 0736+053, WD 0743-366, WD 1009-184, WD 1132-325, WD 1544-377, WD 1620-391

The above discussion strongly suggests there are many SLSs beyond 20 pc remaining to be found. Why have they not yet been identified? Of the 11 systems within 20 pc, six are wide binaries where the WD is not obscured by the glare of the main sequence star. It can be argued that efforts to identify low luminosity common proper motion (CPM) companions of nearby bright stars, such as that of Gould & Chanamé (2004) and Tokovinin & Lépine (2012), who searched around *Hipparcos* stars, would have revealed most of the nearby CPM WDs. Indeed some were found and are included in Table 1. However, it is still possible additional wide SLSs will eventually be discovered within 20 pc. For example the system WD 1009-184 + BD -17° 3088, at a distance of 17 pc and a separation of 400″, was accidently discovered during the compilation of the revised 20 pc WD sample (Holberg et al. 2008). It is quite possible that more such widely separated SLSs remain within 25 or 30 pc of the sun. The remaining five SLSs within the 20 pc local volume are near enough to their MS partner that they are difficult to either detect or to obtain spectra confirming their WD status. For example, the WD in GJ 86 (WD 0210-508) was suspected of being a cool DA, however, recent *HST* spectra (Farihi et al. 2013) have revealed it to be a DQ star. Unless such systems reveal themselves through astrometric perturbations or radial velocity variations or perhaps are found through adaptive optics imaging of the vicinity of bright stars they are likely to remain overlooked. Indeed the WD nature of both GJ 86 B and ε Ret B (WD 0416-593) were recognized by attempts to image exoplanets near the primary.

## 5 FINDING ADDITIONAL SIRIUS-LIKE SYSTEMS

### 5.1 The Gaia mission

*Gaia* is an ambitious astrometric mission planned for launch by the *European Space Agency (ESA)* in late 2013, as a follow-up to the *Hipparcos* mission. An excellent summary of the general capabilities of *Gaia* mission is given in de Bruijne (2012) and with respect to WDs in particular by Garcia-Berro et al. (2005). It will operate as an all-sky survey producing positions, proper motions, parallaxes, photometry and radial velocities for up to one billion stars. For a $15^{th}$ magnitude star at the end of the nominal five-year mission, positional accuracies of 25 µas and proper motion accuracies of 13 µas yr$^{-1}$, or better (depending on V- I color) are anticipated. In general, *Gaia* will yield WD parallaxes some 20 to 50 times more precise than most existing ground-based and *Hipparcos* parallaxes, thus greatly improving distance estimates for most SLSs discussed in this paper. Moreover, during the lifetime of the *Gaia* mission orbital motions for bright stars exceeding 5 µas yr$^{-1}$ are potentially detectable and thus could be used to measure orbital motion in wide binaries. This aspect will be discussed in more detail below. It should be kept in mind that *Gaia* has an expected bright limit of $6^{th}$ magnitude and that stars between $6^{th}$ and $12^{th}$ magnitude will be observed with shortened exposure times (de Bruijne 2012). Thus, some of the luminous stars in Table 2 cannot be directly observed by *Gaia*, however, the WDs in resolved systems can be observed.

The parallax precisions will be a cumulative product of the *Gaia* mission, and most are expected to become available in final form some eight years after launch. Effectively *Gaia* will remove parallax distances as a significant uncertainty in radius determinations for WDs out to a distance of at least 100 pc (Holberg, Oswalt & Barstow 2012). Radii uncertainties will then be dominated by photometric and/or surface gravity and the gravitational redshift uncertainties.

From the results in Table 2 for the average angular motions (Column 7) it is evident that *Gaia* will easily observe the mutual arcs of orbital motion for both components of all known sufficiently resolved systems, even the most widely separated pairs with the longest periods. For example, the wide binary containing WD 1620-391 has a projected orbital separation of 4410 AU and estimated period of 263,000 yrs and the corresponding estimated relative angular motion of 5.2 mas yr$^{-1}$.

With respect to *Gaia* astrometry it is useful to consider the potential for the discovery of additional SLSs. The most easily detected are wide binary CPM pairs. *Gaia's* precise global astrometric capabilities will identify any remaining undetected systems at separations from tens to hundreds of arc seconds based on common proper motions alone; however, precise trigonometric parallaxes at the 50 µas level will also provide common distances as an additional identifier of CPM systems. Likewise, *Gaia's* dispersive photometric system will permit the identification of WDs and even estimates of WD spectral types (Garcia-Berro et al. 2005). Of more specific interest will be *Gaia's* ability to identify potential unresolved systems through recognition of non-rectilinear motions of the luminous star, in much the same way that the existence of the companions of Sirius and Procyon were first discovered.

As can be seen in Table 2, astrometric proper motion accelerations of the luminous companions of many systems will be detectable by *Gaia* during its nominal five-year mission lifetime. Generously assuming that a proper motion acceleration of 8 µas yr$^{-2}$ is measurable over 5 years the presence of possible companions will be evident for a number currently unknown SLSs with *Gaia*. For example from Table 2 the proper motion accelerations of HD 15638 (WD 0226-615), which was unresolved with *HST*, and HD 27843 (WD 0418+137), which was resolved, and has an estimated 348 yr. orbital period, will both be easily measurable with *Gaia*. The point of this discussion is that other potential SLSs are likely to be discovered by *Gaia*. A broader question, however, is to determine if it will be possible to confirm the WD nature of a companion that remains undetected spectroscopically or photometrically. This will require follow-up observations of the type described below.

## 5.2 Ongoing radial velocity surveys

At some level unresolved SLSs must be radial velocity variables. Indeed systems like WD 0457-103 (Queloz et al. 2000), and WD 2124+191 (Vennes, Christian & Thorstensen 1998) were known as radial velocity variables prior to their recognition as UV and EUV excess sources (Vennes et al. 1998). Since the advent of intensive radial velocity searches for exoplanets, hundreds of bright main sequence stars have been regularly examined, and from the estimated frequency of SLSs among main sequence stars in this paper, several of these could well contain the radial velocity signatures of long period WDs. Indeed a long term radial velocity drift of 131 m s$^{-1}$ yr$^{-1}$ was removed by Queloz et al. (2000) to determine the radial velocity signature of the exoplanet GJ86b, a value in reasonable agreement with our simplified estimate of 132 m s$^{-1}$ yr$^{-1}$ in Table 2. A survey of Table 2 indicates that present radial velocity surveys could detect the reflex motion due to the WD in nearly two dozen SLSs, including a number where such variations are already known. Interestingly, at least five of the SLSs in Table 1 also contain known or suspected exoplanets or brown dwarfs (WD 0201+086, WD 0347+171, WD 0416-593, WD 1620-391 and WD 0433+270). These are discussed in Appendix A.

## 5.3 Imaging surveys of nearby stars for exoplanets

A technique which has recently identified two nearby SLSs (WD 0210-583 and WD 0416-593) is direct Near IR imaging of main sequence stars in an effort to identify exoplanets. Such searches can involve adaptive optics and coronagraphic techniques in an effort to image regions from a few to a few hundred AU from bright stars. Two comprehensive exoplanet searches of this nature that are planned are the Gemini Planet Imager (GPI; Graham et al. 2007) and Spectro-Polarimetric High-contrast Exoplanet Imager (SPHERE)[3]. Previously unresolved WDs that may be found in such surveys are expected to be some 5 to 8 magnitudes fainter than the primaries, and thus could be confused with M-stars. However, they should be distinguishable by multiband photometry.

## 5.4 *GALEX*

A very large potential source of additional SLSs is the *Galaxy Evolution Explorer* (*GALEX;* Bianchi, et al. (2010), which has imaged a large fraction of the sky in two bands: the far-UV (FUV, 1344 – 1786 Å) and near-UV (NUV, 1771-2831 Å). Effectively, suitable combinations of SLSs with WDs hotter than ~13,000K and companions later than early-F should show significant UV excesses in the *GALEX* bands. Indeed among the stars in Table 1 all of the very close and unresolved systems with suitable later type companions (that have useful *GALEX* data) can be recognized as UV excesses.

## 6 CONCLUSIONS

The SLSs are an important component of our galactic WD population, yet evidence presented here indicates that current surveys, observations and studies have systematically missed large fractions of them, particularly beyond 20 pc. In the past, SLSs were largely confined to resolved CPM binary pairs. But during the last thirty years observations from space have revealed a large number of unresolved, or very difficult to resolve, systems via UV and EUV excesses. Although these latter systems now account for over 30 per cent of the known SLSs, they are largely biased towards very hot H-Rich WD stars. The space density of SLSs is estimated to be 3.3 x10$^{-4}$ pc$^{-3}$ or about 8 per cent of all white dwarfs based on the number currently known within 20 pc. The frequency of SLSs among main sequence stars, from A to K, is estimated to be approximately 1:80 to 1:175. This paper discusses a number of ongoing and new means for detecting possible new SLSs that are likely to become important over the next decade. Both of these estimates are at odds with the number of known SLS beyond 20 pc, indicating many more remain to be discovered.

SLSs are very useful in the study of both the WD Mass-Radius Relations (Holberg et al. 2012) and the Initial-Final Mass Relations. The continued study of these systems will also benefit from improved photometry, surface gravities and gravitational redshifts of the white dwarf components, both from the ground and from space. A number of ongoing and new means of detecting new SLSs will become important over the next decade. These developments should reduce the deficit of such systems beyond 20 pc.

**ACKNOWLEDGMENTS**

---

[3] http://obswww.unige.ch/~wildif/publications/7017-20.pdf

We wish to thank Stefan Jordan for valuable discussions regarding the capabilities of the *Gaia* Mission and Jay Farihi for pointing out corrections to Table 1 and Daryl Willmarth for providing information on the orbit of WD 1109-225 in advance of publication.. J.B.H. acknowledges support from NSF grant AST-1008845 and NASA Astrophysics Data Program grant NNX1OAD76. T.D.O. acknowledges support from NSF grant AST-087919. M.A.B. and M.R.B. acknowledge the support of the Science and Technology Facilities Council, UK. This research has made use of the *WD Catalog* maintained at Villanova University and the *SIMBAD* database, operated at CDS, Strasbourg, France.

**APPENDIX A:** *Notes on Systems*

Below useful information is provided as to the source(s) of the observations used to establish the Sirius-Like nature of each system. Also provided are references to the WD physical parameters and oth3r3 data given in Tables 1 and 2.

**WD 0022-745:** Identified by Burleigh, Barstow, & Fleming (1997) as a *ROSAT* EUV excess source associated with HD 2133. The WD mass is estimated from the $T_{eff}$ and log g of Burleigh et al. (1997). This close DA + F7V system was resolved with WFPC2 by B01.

**WD 0023-109:** Wide CPM companion to G 158-77 (Eggen & Greenstein 1965). The WD parameters are from Gianninas, Bergeron, & Ruiz (2011, hereafter GBR). Farihi (2009) suggests that this is a triple system containing an unresolved double degenerate system and also identifies G 158-77 as an M0V on the basis of photometry. However, Gracés, Catalán & Ribas (2011) give a spectral type of K6V, while Arazimová, Kawka & Vennes (2010) give K7.5V. Our mass estimate is closer to K6V.

**WD 0030+444:** Suggested as a CPM by Greenstein (1974). WD parameters are from GBR.

**WD 0040-220:** A wide binary first noted by Greenstein (1984) who gives DA3 + sdM. Our WD parameters are taken from Limoges & Bergeron (2010). Dhital et al. (2010) in the Sloan Low-mass Wide Pairs of Kinematically Equivalent Stars Catalog estimates type and distance of the main sequence companion to be K7.1V. The WD is SDSS J0043302.80-214512.8 and the K star is SDSS J004301.59-214527.9.

**WD 0041+092:** An EUV excess source associated with BD +08° 102, first noted by Fleming et al. (1991). The WD parameters are taken from Kawka & Vennes (2010). The system was unresolved by WFPC2 (B01).

**WD 0042+140:** A CPM system (OHL), listed as G6 + DC. WD possibly shows Ca H & K lines, which would indicate a DZ. Little additional information is available. The V-band magnitude is estimated from the SDSS magnitude.

**WD 0052+115:** A faint CPM system from OS94 listed as DB + late F or early G. No reliable distance estimate is available.

**WD 0114+027:** Unresolved DA companion to the photospherically spotted G5 giant AY Cet. The orbital period of 56.8 days for this star are taken from Simon, Fekel & Gibson (1985), who also estimate a Teff = 18,000 K and log g =8. From the *IUE* Lyman alpha profile we estimate a mass of 0.62 $M_\odot$ for the WD and use the mass estimate of the G5 III companion of 2.2 $M_\odot$ from Tautvaišienē et al. (2011). From their analysis of the radial velocity variations Simon et al. (1985) estimate an orbital *asini* = 0.041AU and give a mass function. Using the mass estimates in Table 2 we find a = 0.041 AU.

**WD 0208-510:** (GJ 86B). The primary K0V star (GJ 86) hosts the giant exoplanet (GJ86b). Searches for further faint companions (Els et al. 2001) detected a faint star 2″ from the primary. Mugrauer & Neuhauser (2005) obtained J, H & K band photometry and showing it to be a WD and presented evidence of orbital motion. Recent *HST* photometry and spectrometry of Farihi et al. (2013) clearly show the orbital motion of the WD as well as identifying the star as a DQ making the system remarkable similar to Procyon A and B. WD parameters, orbital positions and WD mass are from Farihi et al. (2013).

**WD 0210+086:** Unresolved companion to the barium star HD 13611. It was discovered as an UV excess by Böhm-Vitense, & Johnson (1985). The masses of the WD and primary star are poorly known but are taken from Böhm-Vitense, & Johnson. Speckle observations of Roberts (2011) give a separation upper limit of 0.16″ while *Hipparcos* astrometry gives an orbital period of P = 4.496 yrs (Pourbaix, & Boffin 2003).

**WD 0221+399:** A CPM system (OHL) classification spectra gives dK + cool DC. It is also listed as DC in Reid (1996). The WD parameters are from GBR.

**WD 0226-615:** Discovered as an UV excess source associated with the F3IV/V star HD 15638 Landsman, Simon, & Bergeron (1993) and an *ROSAT* EUV excess source by Barstow et al. (1994). The WD parameters used here are taken from Kawka & Vennes (2010). Vennes et al. (1998) found no velocity variations in the primary star. The system was unresolved by WFP2 (B01). The estimated orbital period is less than 54 yrs.

**WD 0227-005:** A CPM pair from OS94. WD parameters are taken from Eisenstein et al. (2006), Kleinman et al. (2013) however, give no mass estimate. Our distance is estimated spectroscopically from the K0V companion.

**WD 0250-007:** A CPM system (OHL). WD parameters are from GBR.

**WD 0252-055:** The EUV excess source associated with HD 18131 discussed by Vennes et al. (1995). The WD parameters used here are from Kawka & Vennes (2010). Vennes et al. (1998) noted evidence of low level radial velocity variations. The system was unresolved by WFPC2 (B01).

**WD0304+154:** A narrowly separated CPM system from OS94, which list the system as DA + K6V. Lee (1984) gives K3V and a spectroscopic parallax of 0.014".

**WD 0315-011:** A CPM system (OHL). WD data is from Catalán et al. (2008b).

**WD 0338-388:** Identified as a magnetic DB (Riemers, et al. 1998). However, Schmidt et al. (2001) interpret the spectra as an unresolved hot non-magnetic DA plus a K5V companion, of comparable magnitude.

**WD 0347+171:** This is the well studied eclipsing pre-CV system V 471 Tau. WD and companion parameters are taken from O'Brian, Bond & Sion (2001). Guinan & Ribas (2001) have used eclipse timings to infer a third body, possibly a brown dwarf, in a 30 year orbit.

**WD 0353+284:** An EUV source and a barium-rich K2V star (Jeffries & Smalley 1996).

**WD 0354-368:** First detected as the X-ray source MS0354.6-3650 by *Einstein* (Stocke et al. 1991) and later as the EUV source EUVE J0356-366 by Christian et al. (1996) from which the WD parameters were taken. The system was resolved by WFPC2 (B01).

**WD 0400-346:** The WD was identified as a CPM companion separated by some 64″ from HD 25535 by Gould & Chanamé (2004). The WD parameters are taken from Kawka, Vennes & Németh (2010).

**WD 0413-077 (40 Eri B):** This is a well known triple system with the DA white dwarf and companion M star (40 Eri C) orbiting the K0.5V star HD 25535 at distance 415 AU. While the dynamical mass and gravitational redshift for the white dwarf are known with respect to 40 Eri C (see Holberg et al. 2012), the Sirius-Like nature of this system pertains top 40 Eri A.

**WD 0415-594 (ε Ret B):** Long known as a resolved binary system, the faint companion was not recognized as a WD until Chauvin et al. (2006) obtained near IR photometry demonstrating its degenerate nature. Echelle spectra of the WD obtained by Farihi et al. (2011) revealed it to be DA3.5 WD in a P >2665 yr orbit. This system also contains a giant planet (HD 27442b) with a mass > 1.6 $M_J$ in a 428 day orbit.

**WD 0418+137:** The WD companion of the Hyades Cluster star HD 27483 was discovered from an *IUE* UV excess by Böhm-Vitense et al. (1993). The primary consists of two F6V stars in a 3.05 day orbit. In estimating the orbital period of the WD, the primary mass in Table 2 assumes two F6V stars. This system was resolved at 1.276" by B01. We estimate orbital period of 275 yrs, well above the system scenarios considered by Böhm-Vitense et al.

**WD 0421+338:** First discovered by Landsman, Simon, & Bergeron (1996) as a UV excess in 56 Per. The WD parameters are taken from Landsman et al. (1996). 56 Per is quadruple system resolved by WFPC2 (B01), and is discussed in the text.

**WD 0433+270:** A cool CPM companion to HD 283750 (OHL). It also contains a brown dwarf candidate HD 283750b system with a separation of 0.025″ (Lucas & Roach 2002). WD parameters are taken from Zhao et al. (2011).

**WD 0457-103:** Noted as an *IUE* UV excess source by Landsman et al. (1993) and an EUV excess source by Wonnacott, Kellett & Strickland (1993). The primary is the K0IV star HR 1608. Vennes et al. determine a radial velocity period of 903 days. The system was not resolved by B01. The WD parameters are taken from Kawka & Vennes (2010).

**WD 0458-364:** This corresponds to the EUV and UV excess in the uncataloged V= 13.7 K6V star discussed by Burleigh, Barstow & Holberg (1998).

**WD 0512+326:** Hodgkin et al. (1993) first noted the EUV excess in the C component of the 14 Aur system. B01 have shown this system to consist of at least five physical components: an unresolved spectroscopic binary 14 Aur A (KW Aur) of spectral type A9V; an unresolved spectroscopic binary 14 Aur Ca of spectral type F2V; and 14 Aur Cb which is the DA WD responsible for the EUV excess. The nature of short period (days) for the orbital companions of 14 Aur A and 14 Aur C is unknown. In Table 1 the separation and position angle of the WD are with respect to Ca, Component A is more distant at 15″. Vennes et al. (1998) find an orbital period of 0.00819 yrs, however as noted by B01 this cannot be due to the WD. We estimate an orbital period in Table 2 of 2430 yrs. The WD parameters used here are those of Vennes et al. (1998).

**WD 0551+123:** A narrowly separated CPM system from OS94, who give spectral types of DA + G9V.

**WD 0551+560:** A narrowly separated CPM system from OS94. Very little is known concerning either component.

**WD 0615-591:** A CPM system from OHL. The WD parameters are from Bergeron et al. (2011).

**WD 0642-166:** Well known resolved companion to Sirius with a well established orbit and within 20 pc.

**WD 0642-285:** A CPM system from OS94. WD parameters are from GBR and the WD V magnitude from T. Oswalt & J.A. Smith, private communication.

**WD 0658+712:** A widely separated CPM pair noted by Greenstein (1980) who designated the WD a DC. Little is known of the faint DC WD. Our V magnitude comes from T. Oswalt & J.A. Smith, private communication.

**WD 0659+130:** Noted as an unresolved EUV source by Vennes et al. (1997). The Distance of 115 pc is from Barstow et al. (2010), while WD parameters are taken from Kawka & Vennes (2010).

**WD 0727-387:** The unresolved hot DA companion to the B3V star HD 59635 (y Pup) detected as an EUV excess source by Burleigh & Barstow (1998). Parthasarathy et al. (2007) estimate the mass of the white dwarf to be 1 $M_\odot$ based on the assumption that it is the remnant of a star more massive than HD 59635. Vennes (2000) has shown that HD 59635 is actually a close binary with a B6 V companion in a 15 day orbit.

**WD 0736+053:**   The well studied resolved companion to Procyon with a well established orbit. The WD and K5IV/V companion parameters are taken from Liebert et al. (2013).
**WD 0743-336:**   VB 03 is the CPM companion of HR 3018 (Van Biesbroeck 1961) and one of coolest WD stars and widest binary among the Sirius-Like systems (Kunkel, Liebert & Boronson 1984).
**WD 0834+576**: Little is known of this system beyond its listing in Stepanian (2005) as 'DA +G'. SDSS photometry shows composite colors. In the *GALEX* survey this system also appears as a UV excess object.
**WD 0842+490**: This DA1.3 WD was discovered as a suspected CPM companion to the A2V star HD 74389 by Sanduleak & Pesh (1990). Its separation is (20.1″) and its 84.3° position angle places it inconveniently near the diffraction spikes from the primary star in many telescopes (Liebert, Bergeron & Saffer 1990). In support of the interpretation of a common distance for the WD and the A2 star Holberg, Bergeron, & Gianninas (2008) find a spectroscopic distance of 119 ± 2.25 pc vs 111.5 ± 7 pc for HD 74389 from the *Hipparcos* parallax. Also evident in 2MASS and SDSS finder images is a second bluish star with a separation and position angle of 10″ and 345°. Recent Hα spectra of this star show a weak narrow Hα absorption profile embedded in a featureless continuum, reminiscent of sdB stars. This star appears to share the proper motion of HD 74389. Not previously reported, is a third star buried within the glare of the primary at an estimated separation and position angle of 5″ and 270°. Recent spectra of this star show it to be an M-star; no proper motion information is available.
**WD 0845-188**: Noted as a CPM system by Greenstein (1984). The WD parameters are from Bergeron et al. (2011). The WD is a flux standard.
**WD 0905-724:**   First detected as a UV excess source associated HR 3643 by Landsman et al. (1996), where the WD parameters used here are taken.   The system was not resolved with WFPC2 by B01.
**WD 0911+023:**   Is the unresolved hot DA companion to the B9.5V star θ Hyd discovered by Burleigh & Barstow (1999).
**WD 0930+815:**   Discovered as a UV excess in the K3 giant HD 81817 by Reimers (1984).
**WD 1004+665:**   A wide pair from OS94. Distance of 63 pc estimated from Reimers (1984).
**WD 1009-184:**   The CPM companion to BD-17 3088 noted by Holberg et al. (2008) and within 20 pc.
**WD 1021+266:**   Associated with the EUV excess in HD 90052 (Vennes et al. 1998). The system was not resolved by B01.
**WD 1024+326:** Identified as an *EUVE* excess source by Génova et al. (1995). The WD parameters are taken from Kawka & Vennes (2010).
**WD 1027-039:**   A wide pair from OS94.  Distance estimate from Zhao et al. (2011).
**WD 1107-257:**   A wide pair from OS94 and also from Gould & Chanamé (2004).
**WD 1109-225:**   The EUV excess in β Crt was discussed in Fleming et al. (1991). Burleigh et al. (2001) analyzed the *FUSE* and *EUVE* spectra of β Crt, finding a very low mass (0.43 $M_\odot$) indicating a history of interacting binary evolution resulting in a He-core WD. The system was not resolved by B01, indicating a semi-major axis of less than 9.3 AU and an orbital period of less than a decade. There is ample evidence of radial velocity variations but no agreement on an orbital period. Recently a spectroscopic orbit for the primary shows a ∼ 6 yr orbital period, (D. Willmarth, private com.).
**WD 1130+189:**   A wide pair from OS94, which list the system DA + G2V.
**WD 1132-325:**   VB 04 is the CPM of HD 100623 (Van Biesbroeck 1961). Although separated by 16″ from its 6$^{th}$ mag. primary star there is no definitive determination of temperature or mass. A spectrum showing its DC nature is shown in Giammichele, Bergeron & Dufour (2012). Dieterich et al. (2012) gives K0.0V for he primary.
**WD 1133+619:**   A wide pair from OS94 and also from Gould & Chanamé (2004). OS94 lists the system as DZ + K5V.
**WD 1209-060:**   A wide pair from OS94, which lists the system as DC + K5V. WD parameters from Garcés et al. (2008)
**WD 1227+292:**   A wide pair from OHL, little else is known about this system.
**WD 1250-226:**   A barely resolved hot WD around the central star of the planetary nebula PK 303+40°.1 (A35). The WD parameters are from Heald & Bianchi (2002) who studied this system with, *FUSE*, *STIS* and *IUE*. The separation is the mean of the values given in Gatti et al. (1998).
**WD 1304+227:**   A wide pair from OS94. The WD parameters are from GBR.
**WD 1306+083:**   This source corresponds to the EUV source RE J1309+081 and it is thought to be hot DA WD companion to a V = 11.5 F9V star. The SDSS field shows no blue objects other than the F9V star. B01 failed to resolve this source with WFPC2 imaging.
**WD 1354+340:**   A wide pair from OS94. The WD parameters are from GBR.
**WD 1425+540:**   Noted as a CPM by Greenstein (1984). The WD parameters are from Bergeron et al. (2011).
**WD 1440+068:**   A wide pair from OS94.
**WD 1455+300:**   Identified by Gould & Chanamé (2004) as a CPM companion to BD +30° 2592 with a separation of 23″.
**WD 1501+301:**   A wide pair from OS94. SDSS spectrum shows a featureless DC continuum. Assuming a log g of 8 the SDSS photometry indicates a $T_{eff}$ of 7000 K a distance of 79 pc and a mass of 0.79 $M_\odot$.
**WD 1514+411:**   A wide pair from OS94. The DA WD was observed by Eisenstein et al. (2006) who obtained $T_{eff}$ = 12,960 ± 140 K and log g = 7.86 ± 0.03. The distance of 139 ± 3 pc is calculated from the SDSS photometry.
**WD 1516+207:**   This unresolved Sirius-Like system involving a G8II-III barium star was noted as a spectroscopic binary (de Medeiros & Mayor 1999). This was followed by (Frankowski et al. 2007) who subsequently noted it to be a proper motion binary from *Hipparcos* data. Stefanik, et al. (2011) have analyzed all available radial velocity and astrometric data, including an unpublished *IUE* spectrum which clearly establishes the DA nature of the companion.   They have determined an orbital period of 1.71 yrs and estimated the temperature and mass of the WD companion and re-estimated the *Hipparcos* parallax. In Tables 1 and 2 we have used the Stefanik results to estimate a spectroscopic mass of 0.80 $M_\odot$ for the WD.

**WD 1542+729:** A CPM pair from OS94, where the WD is listed as a DC. This system is also noted in Gould & Chanamé (2004), with no WD spectral type given.

**WD 1544-377:** Noted as a CPM companion by Wegner (1973). The WD parameters are from GBR.

**WD 1619+123:** Noted as a CPM by Farihi, Becklin, & Zuckerman (2005). The WD parameters are from GBR.

**WD 1620-391**: Discovered as a nearby wide CPM system by Stephenson, Sanduleak, & Hoffliet (1968). A bright well studied DA WD within 20 pc, whose companion HR 6094 (HD 147513) also has an exoplanet (HD147513b) with a period of 528.4 days at a distance of 1.32 AU. The WD is widely separated (4895 pc) from HR 6094 and its exoplanet. The WD parameters are taken from Holberg et al. (2012).

**WD 1623+022:** (Also LPSM J1626+0210N, from Lépine & Shara 2005) Identified by Gould & Chanamé (2004) as a CPM companion to LSH 3194. Reid & Gizis (2005) give DC spectral type.

**WD 1634-573:** The DOZ1companion to HD149499. It is both a UV and EUV source and the WD parameters are taken from Jordan et al. (1997). The separation and position angle are taken from Holden (1977).

**WD 1635+530:** Is an unresolved companion to the B9.5V star 16 Dra discovered by Burleigh & Barstow (1998). 16 Dra is in turn associated with 17 DraA/B, respectively a B9V and A1V pair. The $T_{eff}$ and log g of the WD are poorly defined by the *EUVE* spectrum.

**WD 1659-531:** A wide pair from OHL. The WD parameters from GBR.

**WD 1706+332:** Noted as a CPM by Greenstein (1984). The WD parameters are from Holberg et al. (2012).

**WD 1710+683:** Noted as a CPM by Greenstein (1984). The WD parameters are from GBR.

**WD 1730-370:** The unresolved companion to λ Sco which Burghoeffer, Vennes & Dupuis (2000) identify as source of the soft X-ray and EUV excess as a massive hot DA WD. The primary has an orbital velocity period of 5.959 days but the companion is uneclipsed. Burghoeffer et al. use EUV and orbit constraints to estimate the companion must be very massive (> 1.25 $M_\odot$ and hot (>64000 K).

**WD 1734+742:** The WD is a companion to the Ba II star (HD 160538). Fekel et al. (1993) find an *IUE* UV continuum corresponding DA with a $T_{eff}$ = 30000 K and log = 8 (which corresponds to a mass of 0.65 $M_\odot$) and a radial velocity orbit having a period of 14.24 yrs and an estimated separation of 1.84 AU.

**WD 1736+133:** Discovered by accidently by Böhm-Vitense (1992) with *IUE* while observing HD 160365. Böhm-Vitense determines that the WD is a visual binary located approximately 8" south of the F6III star.

**WD 1750+098:** Noted as a CPM in Eggen & Greenstein (1965). The WD parameters are from GBR.

**WD 1803-482:** Noted as CPM by Eggen & Greenstein (1965). The WD is a companion to the Ba II star (HD 165141). Fekel et al. (1993) find an *IUE* UV continuum corresponding DA with a $T_{eff}$ = 35000 K, log = 8 (which corresponds to a mass of 0.66 $M_\odot$) and a radial velocity orbit having a period of 2.47 yrs an estimated separation of 0.277 AU.

**WD 1848+688** (LDS 2471B): the CPM companion to BD +68 1027 (Lépine & Shara 2005, and Gould & Chanamé 2004))

**WD 1921-566:** Noted as a EUV excess in Barstow et al. (1994). The WD parameters are taken from Kawka & Vennes (2010). The system was resolved with WFPC2 in B01.

**WD 2048+809:** Noted as a CPM by Greenstein (1984). The WD parameters are from GBR.

**WD 2110+300:** Noted as a possible UV excess in the mild barium star ζ Cyg by Böhm-Vitense (1980) and confirmed by Dominy & Lambert (1983), who estimate a $T_{eff}$ ~12000 K for the WD. Griffin & Keenan (1992) give a radial velocity period of 17.8 yrs. WD was resolved with WFPC2 (B01) at 0.032″ giving projected separation of 1.6 AU.

**WD 2123-226:** Noted as a UV excess in the barium star ζ Cap by Böhm-Vitense (1980), who estimate a $T_{eff}$ ~23000 K for the WD.

**WD 2124+191:** First noted as an EUV excess in the A8V star IK Peg (Wonnacott, Kellett & Strickland 1993), and confirmed as a UV excess by Landsman, Simon, & Bergeron (1993). Vennes et al. (1998) determine an orbital period of 0.05947 yrs. The WD is massive (1.20 $M_\odot$) Vennes & Kawka (2008). The physical separation computed from the orbital period and the assumed masses. The system was not resolved with WFPC2 in B01.

**WD 2129+000:** A wide CPM first noted as containing a WD by Kuiper (1943). The WD parameters are taken from the analysis of DB stars by Bergeron et al. (2011).

**WD 2217+221:** Identified as a CPM in Hintzen (1986), also noted as a CPM pair in Gould & Chanamé (2004). Relatively little is known of either star.

**WD 2253-081:** A wide CPM from OS94. The WD parameters are from GBR.

**WD 2257-073:** Noted as an EUV excess system by Barstow et al. (1994). The WD parameters are taken from Barstow et al. (2010) and Kawka & Vennes (2010).

**WD 2258+406:** First noted as a CPM pair by Greenstein (1969), who designated the companion as 'sdK8'. The pair has been included in several gravitational redshift and IFMR studies but no further spectral type is given for the companion. Using the distance modulus for the WD (2.95) the absolute V magnitude of the companion is +8.62, making it a border line K9 or M0 star and the least luminous companion among the Sirius-Likes in this paper. The WD is a ZZ Ceti variable analyzed in Gianninas et al. (2005), where the WD parameters used here are from, including the distance of 38.9 pc.

**WD 2301+762:** A CPM system from OHL. Spectral is type from Greenstein (1984).

**WD 2304+251:** (56 Peg B) The unresolved companion of the mild Ba II star HD 218356 discovered by Schlinder et al. (1982) as a UV spectral excess. Griffin (2006) finds a low amplitude radial velocity curve with a period of 111.140 days and gives a mass function and an *asini* constraint.
**WD 2350-083:** Noted as a CPM by Greenstein (1980). WD parameters from Zhao et al. (2012)
**WD 2350-706:** Noted as an EUV excess source by Barstow et al. (1994). The system was resolved with WFPC2 (B01). The WD parameters are taken from Kawka & Vennes (2010).

**APPENDIX B:** *Possible Sirius-Like Systems*

Listed below are possible Sirius-like systems which presently lack sufficient photospheric observational confirmation of the existence or nature of the WD companion. Five of these are barium stars and one is a G-band subgiant, HD 165162, from Böhm-Vitense et al. (2000), while six are barium dwarfs, from Gray et al. (2011). All of these show evidence of barium excesses. Provisional WD numbers have been assigned for these systems. Two stars, Regulus and HD 209295 are radial velocity variables with probable degenerate companions. Also of note is the search for wide CPM systems among Hipparcos FGK stars by Tokovinin & Lépine (2012, see Table 2) that identifies 21 probably systems containing WDs out to a distance of 67 pc. Four of these stars ( WD 1107-257, WD 1619+123, WD 2129+000, WD 2217+211) were already included in our survey of SLSs on their own merits. We have not included any of the other Tokovinin & Lépine CPMs due to the lack of spectroscopic identification and the significant probability that many are not physical pairs.

**Table B 1**
**Possible Sirius-Like Systems**

| WD Num. | Name | Sp. Type | V | Dist(pc) | Comment |
|---|---|---|---|---|---|
| '1202+090' | HD 104979 | G8 III | 4.12 | 50.05 | Böhm-Vitense et al. (2000) |
| '1500+405' | HD 133208 | G8 IIIa | 3.49 | 69.06 | Böhm-Vitense et al. (2000) |
| '0630-111' | HD 46407 | G8 III | 4.12 | 123.30 | Böhm-Vitense et al. (2000) |
| '1804-282' | HD 165634 | G5 IV | 4.56 | 114 | Böhm-Vitense et al. (2000) |
| '0025+099' | HD 2454 | F5 V | 6.04 | 36.6 | Gray et al. (2011) |
| '0225+005' | HD 15306 | F4 V | 8.92 | …. | Gray et al. (2011) |
| '0419+584' | HD 26367 | F7 V | 6.56 | 38 | Gray et al. (2011) |
| '0518+640' | HD 34654 | F8.5 V | 7.29 | 60 | Gray et al. (2011) |
| '1308+214' | HD 114520 | F5 IV-V | 6.86 | 141 | Gray et al. (2011) |
| '2330-122' | HD 221531 | F 5V | 8.33 | 96 | Gray et al. (2011) |
| '1005+022 | Regulus | B8 IV | 1.42 | 24.13 | Gies et al. (2008) |
| '2200-647 | HD 209295 | A9 V | 7.32 | 117 | Handler et al.. (2002) |

**Note Added in proof:**
An additional Sirius-like system has recently come to our attention. Crepp et al. (2013, ApJ, 774, in press) report that HD 11417 (G5, d = 26.14 pc) has a close white dwarf companion (sep. = 0.692 mas) that has been detected through radial velocity accelerations and astrometric motions of the primary as well as having been resolved with high-contrast imaging.

Figure 1.  The relative distributions of main sequence components of SLSs compared with the relative frequency of main sequence stars in the vicinity of the Sun (see Phillips et al. 2009).  The space distribution of main sequence stars given in Phillips et al. has been normalized to 78 stars, the total number of Sirius-Like primary stars.

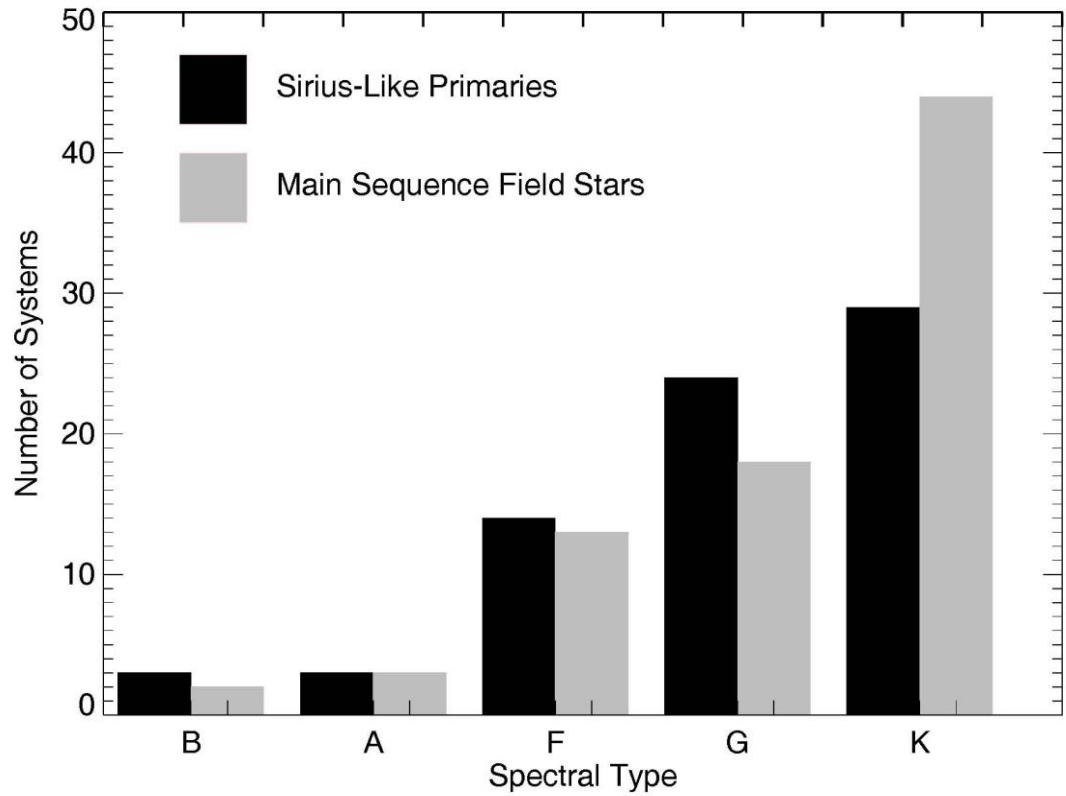

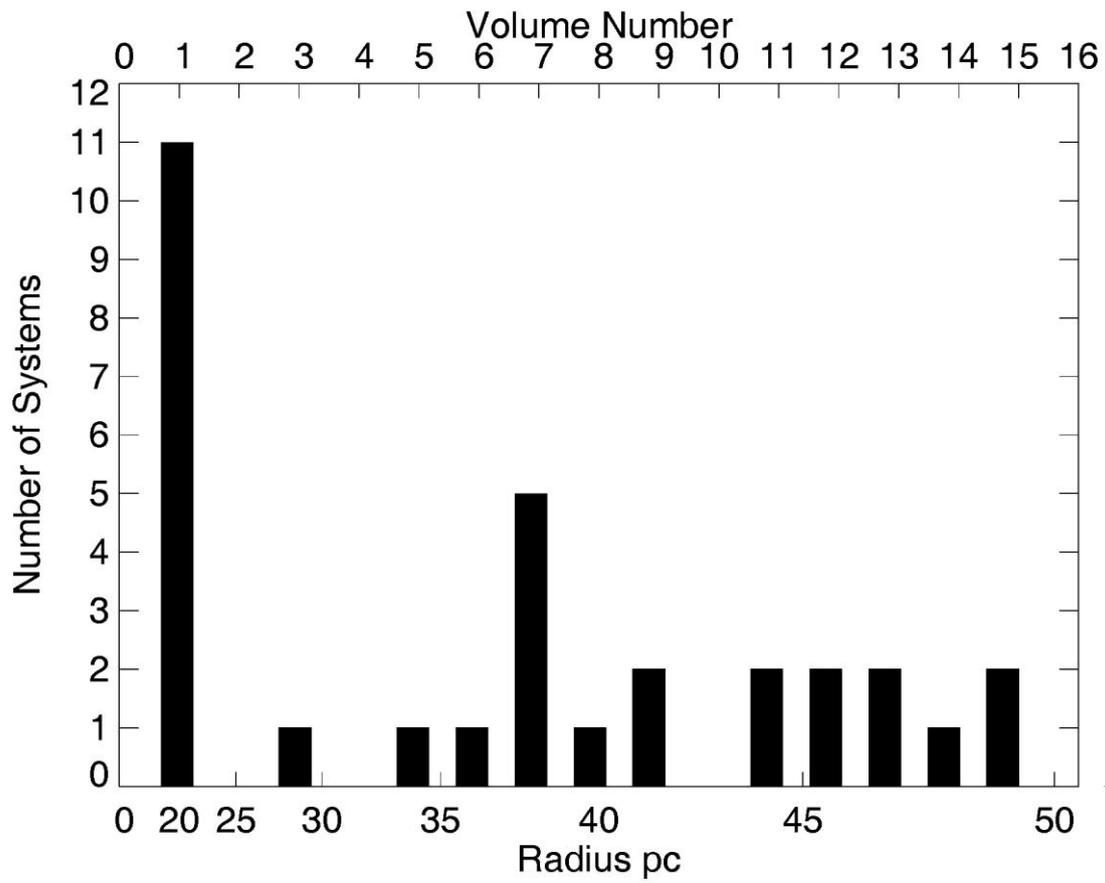

Figure 2. The frequency distribution of SLSs in successive shells of equal volume. The first volume ($3.35\times10^4$ pc$^3$) is the 20 pc local sample where 11 SLSs are known that have a corresponding space density of 3.3 x$10^{-4}$ pc$^{-3}$. Over subsequent shells the mean number of SLSs is 1.2.